\documentclass[preprint2]{aastex6}
\usepackage{multirow}
\usepackage{amsmath}
\usepackage{stmaryrd}
\usepackage{hyperref}
\usepackage{epstopdf}
\usepackage{amsmath,bm}
\usepackage{amssymb}
\usepackage{natbib}
\usepackage{morefloats}
\usepackage{array}
\usepackage{verbatim}

\begin{document}

\title{The scattering of FRBs by the intergalactic medium: variations, strength and dependence on dispersion measures}

\author{Weishan Zhu\altaffilmark{1}, Long-Long Feng\altaffilmark{1,2}, and Fupeng Zhang\altaffilmark{1}}
\affil{$^{1}$ School of Physics and Astronomy,  Sun Yat-Sen University, Zhuhai 519082, China\\
$^{2}$Purple Mountain Observatory, CAS, Nanjing, 210008, China}
\begin{abstract}
The scattering of fast radio bursts (FRBs) by the intergalactic medium (IGM) is explored using cosmological hydrodynamical simulations. We confirm that the scattering by the clumpy IGM has significant line-of-sight variations. We demonstrate that the scattering by the IGM in the voids and walls of the cosmic web is weak, but it can be significantly enhanced by the gas in clusters and filaments. The observed non-monotonic dependence of the FRB widths on the dispersion measures (DM) cannot determine whether the IGM is an important scattering matter or not. The IGM may dominate the scattering of some FRBs, and the host galaxy dominates others. For the former case, the scattering should be primarily caused by the medium in clusters. A mock sample of 500 sources shows that $\tau_{\rm{IGM}} \propto \rm{DM_{IGM}}^{1.6-2.1}$ at $z<1.5$. Assuming that the turbulence follows Kolmogorov scaling, we find that an outer scale of $L_0\sim 5\,$pc is required to make $\tau_{\rm{IGM}} \sim 1-10\,$ms at $\nu=1\, $ GHz. The required $L_0\sim 5\, $pc can alleviate the tension in the timescales of turbulent heating and cooling but is still $\sim 4$ orders of magnitude lower than the presumed injection scale of turbulence in the IGM. The gap is expected to be effectively shortened if the simulation resolution is further increased. The mechanisms that may further reduce the gap are shortly 
discussed. If future observations can justify the role of the IGM in the broadening of FRBs, it can help to probe the gas in clusters and filaments.
 \end{abstract}

\section{Introduction}
Fast radio bursts (FRBs) are a recently discovered class of millisecond-duration radio transients (e.g., \citealt{Lorimer2007, Thornton2013,  Champion2016, Petroff2016}). The dispersion 
measures (DMs) of observed FRBs range from 176 to a few thousands (the highest value until now is 2596 ${\rm{pc\, cm^{-3}}}$; see \citealt{Bhandari2018}). For most observed events, 
their DMs are much higher than the expected value because of the medium in the Milky Way, which indicates that FRBs are of extra-galactic origin. It is suggested that the DMs of FRBs are 
substantially contributed by the intergalactic medium (IGM), which can be used in principle to probe the properties of the IGM(\citealt{Ioka2003, Inoue2004, Deng2014, McQuinn2014}). 

The broadening of the pulse width because of scattering by the turbulent medium, i.e., $\tau$, is also an important parameter of FRBs. The scattering by the Galaxy is inadequate to explain the broadening of several FRBs at high latitudes. The location of the non-galactic scattering has not been determined because both the host galaxy medium and IGM may play important roles. The host can cause significant broadening that is sufficiently strong to explain the observation (\citealt{Cordes2016}; \citealt{Katz2016}; \citealt{Xu2016}, hereafter XZ16). Meanwhile, the contribution of the IGM is under debate. \citet[hereafter MK13]{Macquart2013} estimated that the broadening contributed by the extended, diffuse IGM at $z<3$ was approximately $\tau_{\rm{IGM}} \la 1 \,$ms at $\nu \sim 300$ MHz. They showed that the intervening halo gas and intra-cluster medium (ICM) along the LOS might be capable of producing $\tau_{\rm{IGM}} \sim 5 \,$ms at $\nu \sim 300$ MHz, but they doubted the probability. Two major concerns were recently raised against the IGM as an important scattering matter: (1) To produce $\tau_{\rm{IGM}} \sim 5 \,$ms at $\nu \sim 300$ MHz, the  
outer scale of turbulence with the Kolmogorov spectrum must be $\sim 10^{-2}\,$pc, which appears too small compared to the often presumed injection scale ($\geq 100\,$kpc) and is incompatible 
with the cooling rate of the IGM(\citealt{Luan2014}, XZ16). (2) The non-monotonic dependence of the observed FRB widths on DMs is inconsistent with the expectations for intergalactic 
scattering(\citealt{Katz2016}). 

However, \citet{Yao2017} argued that the broadening of observed FRBs tended to increase with the DM contributed by the IGM. In fact, the strength of this tendency may have been weakened by the large 
lines-of-sight(LOS) variations in the scatter measure (SM) caused by the clumpy IGM (MK13). The anisotropic gravitational collapse makes the initially small-amplitude density fluctuations of cosmic matter form a 
large scale spatial pattern known as the cosmic web, which consists of clusters, filaments, walls and voids (\citealt{Zeldovich1970, Bond1996}). The density perturbation growth in the late nonlinear stages because of 
gravitational instability can be described by a turbulence model (\citealt{Shandarin1989}). In addition, the accretion of matter to collapsed objects is highly anisotropic and non-homogenous. The gas accreted
 into dark matter halos occurs in both hot and cold mode, and contains clumps of various size(\citealt{Dekel2009}). When low mass dark matter halos falling into the gaseous halo of more massive dark matter halos, both thermal and dynamical instabilities, such as the Kelvin-Helmholtz and Rayleigh-Taylor instability, will be triggered and can lead to density fluctuations on scale smaller than the satellite halos(e.g.,  \citealt{Mayer2006, Abramson2011}).  Recent cosmological hydrodynamical simulations without a star formation process found considerable density fluctuations of gas on the resolution scale, i.e., tens of kpc, particularly in clusters and filaments(\citealt{Vazza2010, Zhu2017}). Cosmological hydrodynamical simulations of galaxy formation have shown density and temperature fluctuation on their resolution scale of a few kpc in regions within and outside the dark matter halos (e.g., \citealt{Vogelsberger2012}, \citealt{Nelson2013} ).  
 
 In this work, we probe the scattering of FRBs by the clumpy IGM using cosmological hydrodynamical simulations and revisit the required outer scales of turbulence that can make the IGM an important contributor to the broadening of FRBs. We present the numerical methodology in Section 2. The DM and SM contributed by the IGM residing in the cosmic web are probed in Section 3. We then probe the scattering of FRBs by the IGM in Section 4. In Section 5, we discuss the required outer scales of turbulence to make the IGM play important roles in the broadening of FRBS. Then we summarize our results in Section 6.

\section{methodlogy}

\begin{figure*}[htbp]
\vspace{-1.0cm}
\hspace{2.5cm}
\includegraphics[width=0.48\textwidth,angle=90]{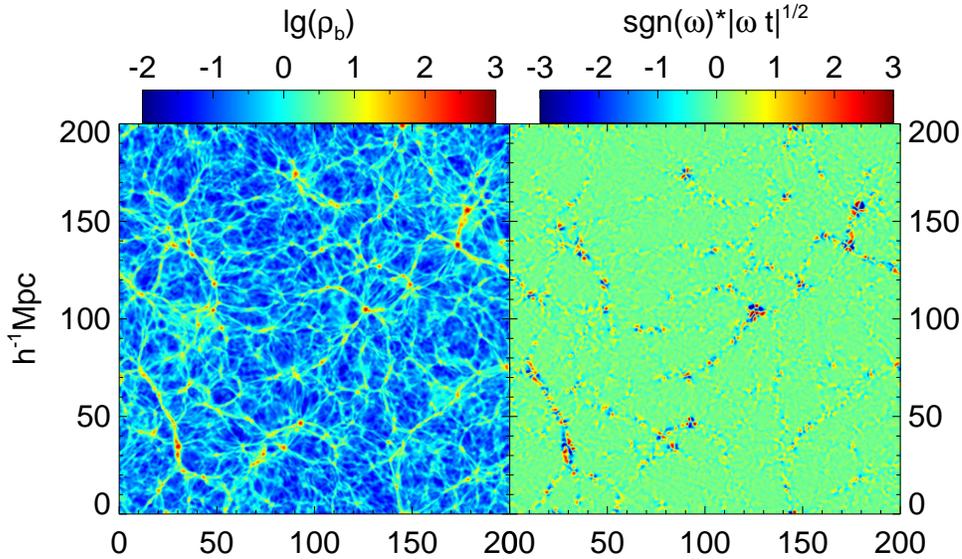}
\vspace{-0.3cm}
\caption{Density (left) and projected vorticity (right) of the IGM in a slice of thickness $0.39 h^{-1}$ Mpc at $z=0$. The red/blue color in the right panel indicates that the vorticity is toward/away from the 
observer.}
\label{figure1}
\end{figure*}

\subsection{simulations}
The IGM distributions in periodical boxes were obtained from a fixed-grid cosmological hydrodynamical simulation using the code WIGEON (\citealt{Feng2004,Zhu2013}, hereafter Z13) with a 
$1024^3$ grid and an equal number of dark matter particles. We expect that a simulation with higher spatial resolution will tend to have larger density fluctuation and scattering measures. To illustrate 
such effects, we ran three simulations with box side lengths of 200, 100 and 50 $ h^{-1}$ Mpc. The corresponding spatial resolutions are 195, 97.7, and 48.8 $h^{-1}$ kpc. The mass resolution of dark matter particles are $7.7\times 10^8, 9.7\times 10^7, 1.2 \times 10^7 h^{-1} M_{\odot}$ respectively. These 
simulations will be referred to as B200, B100 and B050 in the following sections. The Planck cosmology was adopted, i.e., $\Omega_{m}=0.317, \Omega_{\Lambda}=0.683,h=0.671,\sigma_{8}=0.834, 
\Omega_{b}=0.049$, and $n_{s}=0.962$ \citep{Planck2014}. Radiative cooling and heating from a uniform ultraviolet background\citep{Haardt2012} were included. The star formation and active 
galactic nuclei (AGN) were not tracked. At $z\leq2.5$, we successively stored the distribution of the IGM with the redshift intervals given by the light-crossing time through the box. 

Figure 1 shows the density in a slice of thickness $0.39 h^{-1}$ Mpc at $z=0$, which exhibits the cosmic web pattern. The vorticity of velocity $\vec{\omega}=\nabla \times \vec{v}$ is a good indicator of the turbulence in the IGM and ICM (e.g., \citealt{Ryu2008, Zhu2010}). The projected vorticity is also presented in Figure 1, which is rescaled as $sgn(\vec{\omega})|\vec{\omega} t|
^{1/2}$ to increase the contrast, and $t$ is the cosmic time. The turbulence is well developed in the over-dense region, particularly in filaments and clusters/knots. Figure 2 presents the distribution of baryon density in the three simulations mentioned above at $z=1$ and $z=0$, and corresponding cumulative distribution. The number of grid cells that have a baryon density larger than $10$ times of the cosmic mean grows as the redshift decrease. At $z=0$, the volume fraction of cells with $\rho_b/\bar{\rho}_b>100$ in B200, B100 and B050 is about $3.3, 5.9, 7.2\times 10^{-4}$ respectively. The mass fraction of baryons residing in over-dense region with $\rho_b/\bar{\rho}_b>100, 1000$ in B050 is around $25 \%$ and $5 \%$ respectively at $z=0$.

\begin{figure}[htbp]
\vspace{-0.2cm}
\hspace{0.2cm}
\includegraphics[width=0.45\textwidth]{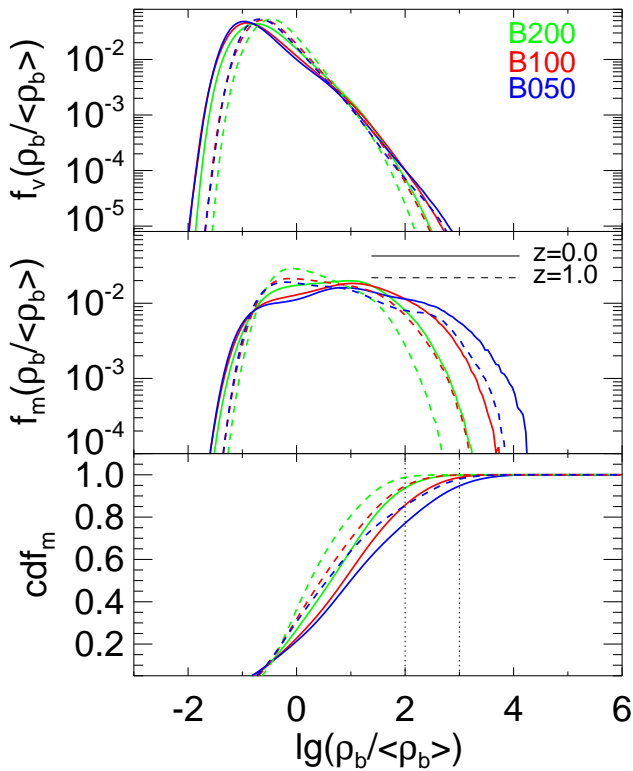}
\vspace{0.8cm}
\caption{Top: The volume fraction of baryonic matter as a function of density contrast in the three simulations mentioned in section 2 at $z=1$ and $z=0$; the bin width is $\Delta[\mathrm{log(\rho_b/<\rho_b>)]=0.05}$, and $<\rho_b>$ indicates the cosmic mean density of baryon. Middle: The mass fraction of baryonic matter in three simulations. Bottom: The cumulative distribution of baryon mass fraction as a function of  density contrast.}
\label{figure2}
\end{figure}

\subsection{Calculation of DM and scattering}
For a source at redshift $z_s$, the dispersion measure caused by the IGM and accounting for the frequency shift due to cosmic expansion is(\citealt{McQuinn2014, Deng2014})
\begin{equation}
\begin{aligned}
{\rm{DM}}(z_s)=\int_{0}^{z_s}\frac{n_e(z)}{1+z}dl,
\end{aligned}
\end{equation}
where $n_e(z)$ is the number density of electrons at redshift $z$. The effective scattering measure because of the extended IGM is given by (e.g., MK13 and XZ16) 

\begin{equation}
\begin{aligned}
{\rm{SM}}_{\rm{eff}}(z_s)=\int_{0}^{z_s}\frac{C_N^2(z)d_H(z)}{(1+z)^3}dz, 
\end{aligned}
\end{equation}
where $d_H(z)=cH_0^{-1}[\Omega_{\Lambda}+\Omega_m(1+z)^3]^{-1/2}$ is the Hubble radius, and $C_N^2(z)$ is related 
to the variance of electron density $\langle\delta n_e^2(z)\rangle$ as
\begin{equation}
C_N^2(z) \approx \frac{\beta-3}{2(2\pi)^{4-\beta}} \langle \delta n_e^2(z) \rangle L_0^{3-\beta}
\end{equation}
when the density power spectrum of the turbulence follows a power law with index $\beta>3$(see XZ16 for $\beta<3$) between 
 the outer scale $L_0$ and the inner scale $l_0$,  assuming $l_0 \ll L_0$. Although there are both supersonic and subsonic turbulence in the IGM (Z13, \citealt{Vazza2017}), 
we only consider the latter and adopt the Kolmogorov turbulence model, i.e., $\beta=11/3$. Following MK13, 
we take $ \langle \delta n_e^2(z) \rangle \sim n_e^2(z)$ , i.e., assuming that the IGM are fully ionized. By and large, the IGM became nearly fully ionized after the reionization 
era, i.e., $z \lesssim 5.5$. A small fraction of IGM remain neutral, which we will not take into account it in this work for the sake of simplicity. Then, the effective scattering measure is

\begin{equation}
\begin{aligned}
{{\rm{SM}}_{\rm{eff}}}(z_s) \approx1.42\times10^{-13} \left(\frac{\Omega_b}{0.049}\right)^2 \left(\frac{L_0}{1 \rm{pc}}\right)^{-2/3}\\
{\rm{m}}^{-20/3} \times \int_{0}^{z_s} [\rho_b(z)/\bar{\rho}_b(z)]^2(1+z)^3d_H(z)dz\\
\approx 1.31\times 10^{13} \  {\rm{m}}^{-17/3} \cdot \frac{1}{h} \  \left(\frac{\Omega_b}{0.049}\right)^2 \left(\frac{L_0}{1 \rm{pc}}\right)^{-2/3}\\
 \times \int_{0}^{z_s} \left[\rho_b(z)/\bar{\rho}_b(z)\right]^2 \frac{(1+z)^3}{[\Omega_{\Lambda}+\Omega_m(1+z)^3]^{1/2}}dz.
\end{aligned}
\end{equation}

Using conventional methods (e.g., \citealt{Gnedin2001,Dolag2015}), we stack the simulation volumes and construct 
light-cones to calculate ${\rm{DM}}(z_s)$ and ${\rm{SM_{eff}}}(z_s)$. 

\section{Contributions to DM and SM by the IGM in the cosmic web }

\begin{figure}[htbp]
\vspace{-0.3cm}
\hspace{0.3cm}
\includegraphics[width=0.45\textwidth]{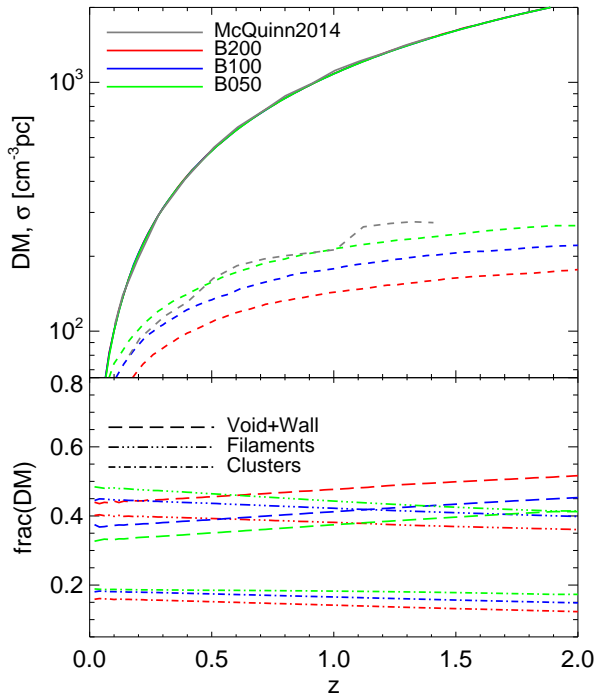}
\vspace{-0.3cm}
\caption{Top: The solid lines indicate the mean values of DM caused by the IGM as a function of 
the redshift, whereas the dashed lines show the standard deviations. The gray lines represent the result in McQuinn 2014. 
The red and blue lines are the results in this work based on simulations B200 and B100, respectively. Bottom: 
The dotted-dashed, triple-dotted-dashed, and long dashed red lines indicate the fractions of DM contributed 
by the gas in clusters, filaments, and voids and walls, respectively.}
\label{figure3}
\end{figure}

\begin{figure}[htbp]
\vspace{-0.3cm}
\hspace{0.3cm}
\includegraphics[width=0.45\textwidth]{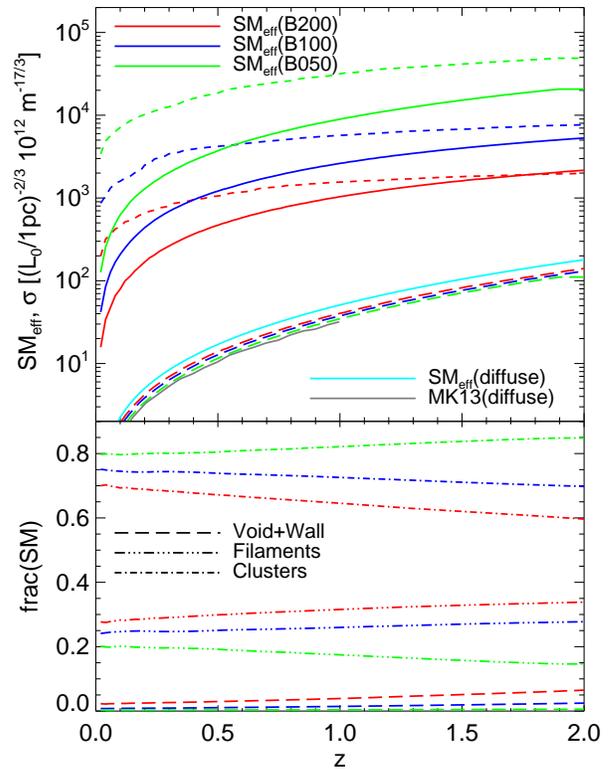}
\vspace{-0.3cm}
\caption{Top: the solid (short dashed) lines indicate the mean values (standard deviations) of ${\rm{SM}_{eff}}(z)$ caused by the IGM as a function of 
the redshift; the blue, red and green lines indicate the results based on simulations B200, B100 and B050, respectively. The long dashed lines show ${\rm{SM}_{eff}}(z)$ caused by gas in voids and walls. 
The gray and cyan solid lines indicate ${\rm{SM}_{eff}}(z)$ caused by the diffuse IGM in MK13 and this work, respectively. Bottom: identical to the bottom panel of Figure 3 but for
 ${\rm{SM}_{eff}}(z)$.}
\label{figure4}
\end{figure}

We randomly sampled 10000 lines of sight for each cosmological simulation. The mean value of ${\rm{DM}}$ contributed by the IGM as a function of redshift $z$, i.e., ${\rm{DM_{IGM}}}(z)$, and the corresponding standard dispersion at $z<2$ are shown by solid and short-dashed lines, respectively, in the top panel of Figure 3. Our results for ${\rm{DM_{IGM}}}(z)$ are consistent with \citet{McQuinn2014} in all three simulations. The deviation in B050 is almost consistent with those of \citet{McQuinn2014}. However, in the other two simulations, the deviations are relatively smaller, which may be a result of the relatively poorer resolution compared to the simulation sample in \citet{McQuinn2014}, which has a mass resolution of dark matter particles of $3\times10^7 h^{-1} M_{\odot}$, and a softening length of 1.6 $h^{-1}$ kpc. Increasing the resolution helps to resolve large density fluctuations in the over-dense region.

The baryonic gas resides in different cosmic structures, i.e., voids, walls, filaments and clusters. To probe the contributions to ${\rm{DM_{IGM}}}(z)$ of different 
structures, the grid cells are assigned to four categories of structures as in \citet{Zhu2017}. As shown in the lower panel of Figure 3, the gas in clusters
contributes approximately $\sim 15\%-20 \%$ of the total ${\rm{DM}_{eff}}(z)$, the gas in filaments contributes $\sim 35-45\%$, and the gas in voids and walls contribute the remaining $\sim 35-45\%$.These fractions basically trace the mass distribution of baryonic matter in the cosmic web as demonstrated in \citet{Zhu2017}.
 
 MK13 analytically estimated the effective SM due to the extended, diffuse IGM by assuming that the IGM had an isotropic and uniform distribution 
 and concluded that it was weak. We run the identical estimation by taking $\rho_b(z)/\bar{\rho}_b(z)=1$ in eqn. (4) and denote it 
 as ${\rm{SM}_{eff, diff}}(z)$. As shown in the top panel of Figure 4, our result for ${\rm{SM}_{eff, diff}}(z)$ is $\sim1.5$ times that in MK13, which 
 should be caused by different cosmological parameters. Then, we derive the mean effective scattering measure as a function of the redshift, i.e., ${\rm{SM}_{eff}}(z)$,
 based on the IGM distribution in our simulations, which is shown as the red, blue and green solid lines in Figure 4. The mean value of ${\rm{SM}_{eff}}(z)$ in B200 is approximately 20 times larger than ${\rm{SM}
_{eff, diff}}(z)$. It is approximately $10^{15} (L_0/\rm{1pc})^{-2/3} \, \rm{m}^{-17/3}$, i.e., $3.25\times 
 10^{-5} (L_0/\rm{1pc})^{-2/3} \, \rm{kpc}\, \rm{m}^{-20/3}$, at $z=1$ in B200. The corresponding ${\rm{SM}_{eff}}$ in B100 and B050 increases by a 
 factor of 2.5 and 10, respectively, compared with B200. For an out scale of $L_0 \sim 100\,$kpc, which was adopted in some previous work(\citealt{Luan2014}, XZ16), ${\rm{SM}_{eff}}$ in B050 is approximately $5\times10^{12} \, \rm{m}^{-17/3}$ at $z=1$.  Apparently, the outer scale influences whether scattering in the IGM/clusters is important or not. We will probe this effect in the next section.
 
The short dashed lines in the top panel of Figure 4 indicate the corresponding dispersions of ${\rm{SM}_{eff}}(z)$. Consistent with the speculation in MK13, ${\rm{SM}_{eff}}(z)$ shows significant
LOS variations with a larger standard deviation than $2$ times the mean value at $z\leq0.5$ in all simulations. We also found that a higher simulation resolution corresponds to 
larger LOS variations in ${\rm{SM}_{eff}}(z)$. These dramatic variations are expected to weaken the dependence of ${\rm{SM}_{\rm{IGM}}}$ on ${\rm{DM}_{\rm{IGM}}}$ if the sample size is small. 
The excess of ${\rm{SM}_{eff}}(z)$ with respect to ${\rm{SM}_{eff, diff}}(z)$ and the dramatic LOS variations should result from 
the intervention of filaments and clusters along some LOS. The long dashed lines in the top panel of Figure 4 indicate that the effective SM caused by gas in voids and walls
is close to the value of ${\rm{SM}_{eff, diff}}(z)$ and contributes only $\sim 4\%-5 \%$ of the total ${\rm{SM}_{eff}}(z)$.
The gas in clusters and filaments contributes approximately $65-80\%$ and $20-30\%$, respectively, as shown in the lower panel of Figure 4. The absolute magnitude of the effective SM contributed 
by the gas in clusters is significantly higher than the estimated value in MK13. Moreover, the contribution by the gas in clusters is relative higher in B050 than in B100 and B200.
The increase in ${\rm{SM}_{eff}}$ in B050 should primarily result from the stronger density fluctuation in the cluster region captured by the increased spatial resolution. As Figure 2 demonstrated,  
there are more cells that have a baryon density $\rho_b/\bar{\rho}_b>10$ in B050 with respect to B200 and B100. According to eqn. 4., the magnitude of ${\rm{SM}_{eff}}$ would be very sensitive to such cells, for a fixed $L_0$. Those cells 
are likely belong to filaments and clusters, considering the density distribution in the cosmic web (\citealt{Zhu2017}).

In short, the gas in clusters contributes only $15\%$ of ${\rm{DM_{IGM}}}$ but dominates the effective SM caused by the IGM.

\section{Time broadening of FRBs by the IGM}

In this section, we study the time broadening of FRBs due to the reported effective scattering measure of the IGM in the last section. For comparison with observations, 
we use the information of the observed FRBs with available scattering times. More specifically, we include 17 events compiled in \citet{Cordes2016} and Y17, 
4 events reported in \citet{Bhandari2018}, event FRB 150807\citep{Ravi2016} and event FRB 170107\citep{Bannister2017}. Many previous theoretical studies investigated the broadening by the IGM at $\sim 300$ MHz (e.g., MK13 and 
\citealt{Luan2014} ). However, the time broadening of the observed FRBs is commonly measured and provided at approximately $\sim 1$ GHz(\citealt{Petroff2016}). 
Thus, for a more accurate comparison between our result and the observational results, we discuss the time broadening at 1 GHz.
\subsection{Variations, dependence on DM, and strength}

The relation between the temporal broadening and $\rm{SM}_{\rm{eff}}$ for Kolmogorov turbulence is (e.g., MK13) 


\begin{equation}
\begin{aligned}
\tau=10^{-4} \  {\rm{ms}} \times (1+z_L)^{-1}\left(\frac{D_{\rm{eff}}}{1\rm{Gpc}}\right) \times \qquad  \qquad  \\
\left\{
\begin{large}
\begin{array}{ll} 
3.32 \times  \left(\frac{\lambda_0}{30 \rm{cm}}\right)^4  \left(\frac{\rm{SM}_{\rm{eff}}}{10^{12} \rm{m}^{-17/3}}\right) \left(\frac{l_0}{1 \rm{AU}}\right)^{-\frac{1}{3}} & r_{\rm{diff}} < l_0   \vspace{3mm}  \\
 9.50 \times \left(\frac{\lambda_0}{30 \rm{cm}}\right)^{\frac{22}{5}}  \left(\frac{\rm{SM}_{\rm{eff}}}{10^{12} \rm{m}^{-17/3}}\right)^{\frac{6}{5}} & r_{\rm{diff}} > l_0 
 \end{array}
 \end{large}
\right.
\end{aligned}
\end{equation}

where $z_L$ is the redshift of the scattering matter; $\lambda_0$ is the wavelength in the observer's frame; 
$D_{\rm{eff}}=D_LD_S/D_{LS}$, with $D_L$ and $D_S$ being the angular diameter distance of the scattering matter and the source from the observer, respectively; and
$D_{LS}$ is the angular diameter distance of the source from the scattering matter. The diffractive length scale $r_{\rm{diff}}$
for $\beta=11/3$ is approximately
 
\begin{subequations}
\begin{equation}
r_{\rm{diff}} \sim (\pi r^2_e \lambda^2_0 \rm{SM}_{\rm{eff}} \mathit{l}_0^{-\frac{1}{3}})^{-1/2}, r_{\rm{diff}}<\mathit{l}_0,
\end{equation}
\begin{equation}
r_{\rm{diff}} \sim (\pi r^2_e \lambda^2_0 \rm{SM}_{\rm{eff}})^{-3/5}, r_{\rm{diff}}>\mathit{l}_0.
\end{equation}
\end{subequations}

As shown in eqns. (4) and (5), $\tau_{\rm{IGM}}$ highly depends on $L_0$ and $l_0$. Although significant 
density fluctuations of gas were found on the resolution scale, i.e., tens of kpc, in many simulations with or without star formation and AGN,
it remains a challenge to probe them by observation. Density fluctuations on similar scales in the central region of the Coma and Perseus clusters were recently reported (\citealt{Churazov2012, 
Zhuravleva2015}). However, constrained by resolution, both simulation and observation cannot provide any information between tens of kpc and the Fresnel scale, i.e., $\sim$ 10 AU. Thus, the value of 
$\tau_{\rm{IGM}}$ discussed here is based on certain assumptions on $L_0$ and $l_0$. 

 \begin{figure}[htbp]
\vspace{-0.0cm}
\hspace{-0.2cm}
\includegraphics[width=0.48\textwidth]{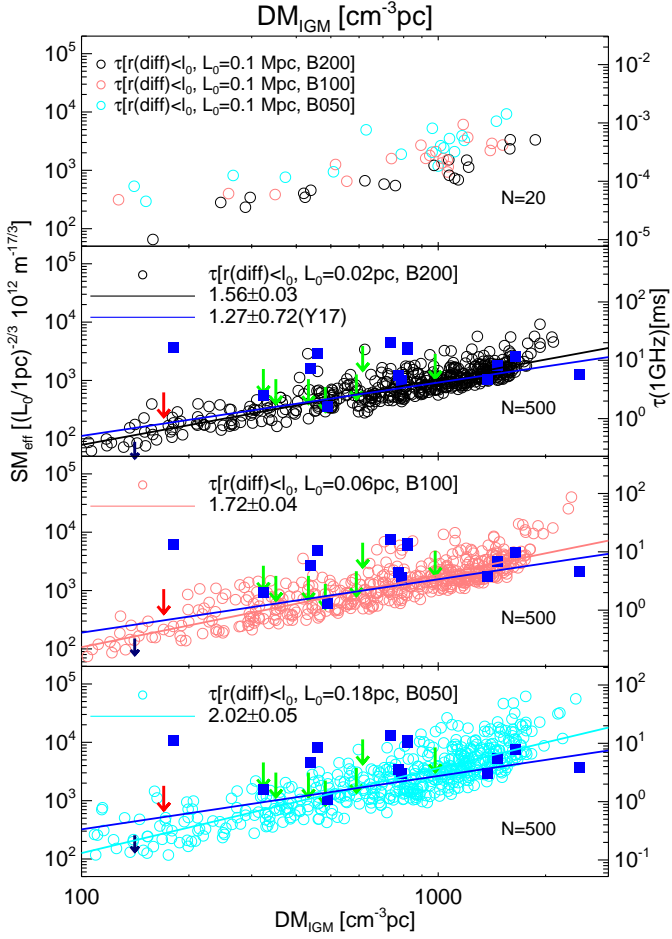}
\vspace{-0.2cm}
\caption{$\rm{SM}_{\rm{eff}}$($\tau_{\rm{IGM}}$) as a function of $\rm{DM_{IGM}}$. 
Top panel: The left axis shows the $\rm{SM}_{\rm{eff}}$ of 20 mock sources, and 
the right axis indicate the corresponding $\tau_{\rm{IGM}}$ at $\nu=1\, $ GHz for the case $r_{\rm{diff}}<l_0$ and $L_0=0.1$Mpc. 
From the second to fourth panels: The circles indicate the $\rm{SM}_{\rm{eff}}$ and $\tau_{\rm{IGM}}$ of 500 mock sources in B200, B100 and B050(the value of $L_0$ is adjusted for each simulation, see 
the text for details). The solid black, pink and cyan lines indicate the linear fitting results of $\tau_{\rm{IGM}}\propto \rm{DM}_{\rm{IGM}}^n$, and the blue solid lines indicate the fitting in \citet{Yao2017}.
The $\tau$ and $\rm{DM}_{\rm{IGM}}$ of the observed events (filled square and downward arrow) are given by data in relevant references(see Petroff 2016). The red arrow 
indicates the repeating FRB 121102. The navy blue arrow indicates the event FRB150807, which has been shifted upward by an order of magnitude.}
\label{figure5}
\end{figure}

For simplicity, we first consider the case $r_{\rm{diff}}<l_0$ with $l_0 \approx 1$AU and assume that
$(1+z_L)\sim1, D_{\rm{eff}}\sim 1 {\rm{Gpc}}$. For the stacked IGM distribution of each cosmological simulation, we randomly place 20 mock sources in the redshift range $0.10<z<1.50$.
The top panel in Figure 5 shows $\rm{SM}_{eff}$ as a function of $\rm{DM}_{\rm{IGM}}$ of these mock sources.
 The low- and high-redshift ends of mock sources are selected so that ${\rm{DM}}$ $(z_{\rm{low}})$ and ${\rm{DM}}$ $(z_{\rm{high}})$ in section 2 are approximately equal to the smallest and largest 
$\rm{DM}_{\rm{IGM}}$ of the observed FRBs. In fact, both host galaxies and the IGM can contribute to the DM of the observed events. Various models have been introduced in the literature, but 
the actual picture remains unclear. So far, it is not straightforward to determine the $\rm{DM}_{\rm{IGM}}$ of observed FRBs using any method. As a beginning, we use a notably simple model 
following \citet{Yao2017}, i.e., assuming that the hosts' contribution to DM is $\rm{DM}_{host}=100 \, \rm{pc\, cm^{-3}}$, except for the repeating event FRB 121102. This simplification can help us to 
reveal the dependence of $\tau_{\rm{IGM}}$ on factors such as $\rm{DM}_{\rm{IGM}}$ and simulation resolution. The host galaxy's contribution for FRB 121102 is set to $\rm{DM}_{host}
=170 \, \rm{pc\, cm^{-3}}$ according to recent observations(\citealt{Spitler2016, Chatterjee2017, Tendulkar2017, Bassa2017}). The assumption of $\rm{DM}_{host}=100 \, \rm{pc\, cm^{-3}}$ is likely 
oversimplified, so we will consider more realistic models in the next subsection.

The dependence of $\tau$ on ${\rm{DM}}$ is used as an important indicator to determine whether the IGM is an important scattering matter of FRBs. For a given $L_0$, 
the dependence of $\tau_{\rm{IGM}}$ on ${\rm{DM_{IGM}}}$ for these 20 mock sources is weak for all three simulations.
The lack of a strong dependence results from the large LOS variations and limited sample size. The variations in 
$\tau_{\rm{IGM}}$ can be up to $\sim 1$ orders of magnitude at $\rm{DM_{IGM}} \sim 1000 \, \rm{pc\, cm^{-3}}$. We then increase the number of randomly distributed mock sources to 500 
for the stacked IGM distribution of each simulation, while the redshift range is kept to $0.10<z<1.50$. The second to fourth panels in Figure 5 show that with a largely increased sample size of 500, $\tau_{\rm{IGM}}$ shows a clear 
positive correlation with ${\rm{DM_{IGM}}}$. The linear least square fitting gives $\tau_{\rm{IGM}} \propto {\rm{DM_{IGM}}}^{1.56\pm0.03}$,  ${\rm{DM_{IGM}}}^{1.72\pm0.04}$,  and ${\rm{DM_{IGM}}}^{2.02\pm0.05}$ in the three simulations.  The linear Pearson correlation coefficient of $\mathrm{log(\tau_{IGM})}$ and $\mathrm{log(DM_{IGM})}$ are $0.90, 0.88, 0.87$ in B200, B100, and B050 respectively. The scaling relation in our work
are steeper than the result in \cite{Yao2017}, i.e., $\tau \propto {\rm{DM_{IGM}}}^{1.27\pm0.72}$, but are within their range of scatter. This discrepancy may be alleviated if the real $\tau$ of those observed FRBs with only upper limits is smaller than the limits. In addition to the large LOS variations of $\rm{SM}_{eff, IGM}$, the dependence of $\tau$ on ${\rm{DM}}$ is also complicated by the mixing contribution to the DM 
by both the host galaxy and IGM. Meanwhile, the selection effect of observed events is not clear. Thus, in contrast to \cite{Katz2016}, the observed non-monotonic dependence 
of the widths of FRBs on DM cannot be used as solid evidence to rule out the IGM considering the limited number of events. 
 
The magnitude of $\tau_{\rm{IGM}}$ is another important criterion to evaluate the contribution of the IGM to the broadening of FRBs.
 According to the estimated value of $\rm{SM}_{\rm{eff}}$ in section 2, 
$\tau_{\rm{IGM}}$ is $\sim 10^{-4}-10^{-3}(10^{-2}-10^{-1}) \,$ ms at the frequency of $1\, $ GHz(300MHz) for ${\rm{DM_{IGM}}} \sim 1000 \, \rm{pc\, cm^{-3}}$,
 with an outer scale of $L_0\sim 100 \ $kpc. This level is implausible for explaining the observed events 
with $\tau \sim 5 \,$ms at $\nu \sim 1$ GHz.  We then adjust the value of $L_0$ in the calculation of $\rm{SM_{eff}}$ for each simulation sample, to make 
$\tau_{\rm{IGM}}{\rm{(1\, GHz)}} \sim 1-10\,$ms for $\rm{DM_{IGM}} \sim 500-1000\, \rm{pc\, cm^{-3}}$. A 
significantly smaller $L_0$ on the order of $\sim 0.02$pc is required in B200, as shown in the second plot of Figure 4. The required
$L_0$ can be increased to $\sim 0.06, 0.18\ $pc in B100 and B050, respectively, because of the stronger density fluctuations captured by the increased resolution. For such values of the outer scale, the 
distribution of $\tau_{\rm{IGM}} - \rm{DM_{IGM}}$ of mock sources can nearly cover that of the observed events, except for FRB010724 and FRB160102. For the former event, we will revisit it 
later. The event FRB 160102 with the highest $\rm{DM}_{xg}$ could be covered by an increased high-redshift end. 

Figure 6 shows the case with $r_{\rm{diff}}>l_0$. For $\rm{DM}_{\rm{IGM}} \sim 500-1000\, \rm{pc\, cm^{-3}}$, the required outer scale $L_0$ to produce $\tau_{\rm{IGM}}{\rm{(1\, GHz)}} \sim 1-10\,$ms 
is $\sim 5\,$pc in B050, which is relatively larger than the case with $r_{\rm{diff}}<l_0$. Meanwhile, convergence 
with increasing resolution is not attained in B050. The demanded outer scale of turbulence is expected to further increase with increasing resolution.

 \begin{figure}[htbp]
\vspace{-0.3cm}
\hspace{-0.2cm}
\includegraphics[width=0.45\textwidth]{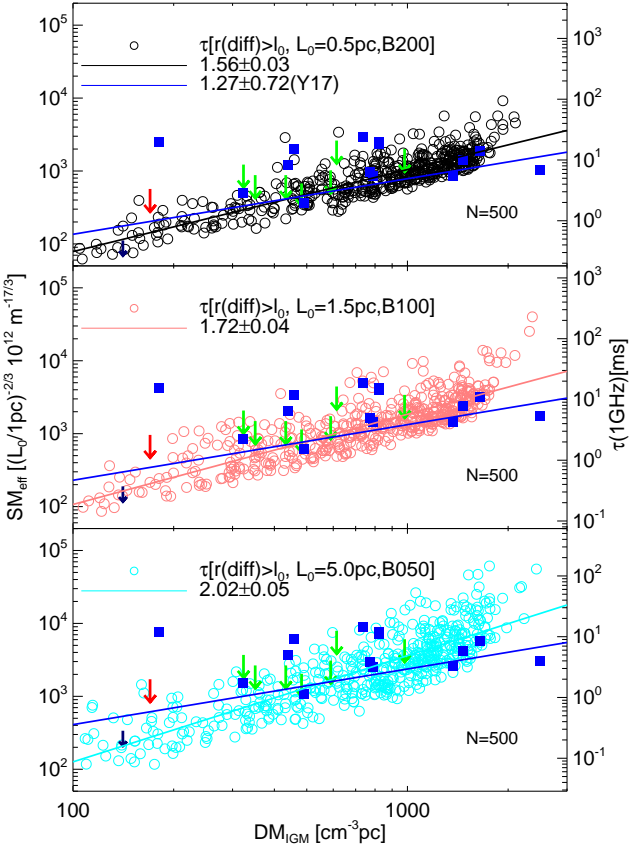}
\vspace{-0.3cm}
\caption{Identical to the bottom three panels in Figure 5 but for $r_{\rm{diff}}>l_0$. }
\vspace{-0.3cm}
\label{figure6}
\end{figure}

\subsection{$\tau_{\rm{IGM}}-DM_{\rm{IGM}}$ relation: the key role of the gas in the clusters}

Assuming that the mock sources are randomly distributed in a certain redshift range, the dependence of $\tau_{\rm{IGM}}$ on $ {\rm{DM_{IGM}}}$ in our simulations 
tends to steepen when the resolution increases. This trend should be related to the enhanced density fluctuations in simulations with higher resolution, particularly in the clusters. We probe this effect by calculating the broadening time due to the effective measure caused by gas in different structures. We use the same mock sources as in the last subsection. In Figure 7, the horizontal axis indicates the dispersion measure contributed by gas in voids and walls, filaments, and clusters. The vertical axes indicate the corresponding effective scattering measure and broadening time, assuming $r_{\rm{diff}}<l_0$. To isolate the impact of resolution on $\rm{SM}_{\rm{eff}}$, as well as the dependence of $\tau_{\rm{IGM}}$ on $ {\rm{DM_{IGM}}}$, we use an identical outer scale of $L_0=0.18\,$ pc for all the samples from three simulations in this subsection.

The top panel shows that the relation between $\tau$ and DM due to the extended, diffuse IGM in voids and walls is consistent in the three simulations, with $\tau_{\rm{void}} \propto {\rm{DM_{IGM,void}}}
^{1.36-1.38}$ with small scatter. Discrepancies appear in the case of gas in filaments, with $\tau_{\rm{filament}} \propto {\rm{DM_{IGM,filament}}}^{1.38}$, $\propto {\rm{DM_{IGM,filament}}}^{1.54}$, 
and $\propto {\rm{DM_{IGM,filament}}}^{1.68}$ in B200, B100 and B050, respectively. Evident dispersion also appears and is enhanced by the increased resolution. The scaling relation between $\tau$ and 
$\rm{DM}$ by the baryonic matter in the clusters is the steepest, with $\tau_{\rm{cluster}} \propto {\rm{DM_{IGM,cluster}}}^{1.64}$, $\propto {\rm{DM_{IGM,cluster}}}^{1.76}$, and $\propto {\rm{DM_{IGM,cluster}}}^{2.00}$ in 
the three simulations.

 \begin{figure}[htbp]
\vspace{-0.3cm}
\hspace{0.0cm}
\includegraphics[width=0.45\textwidth]{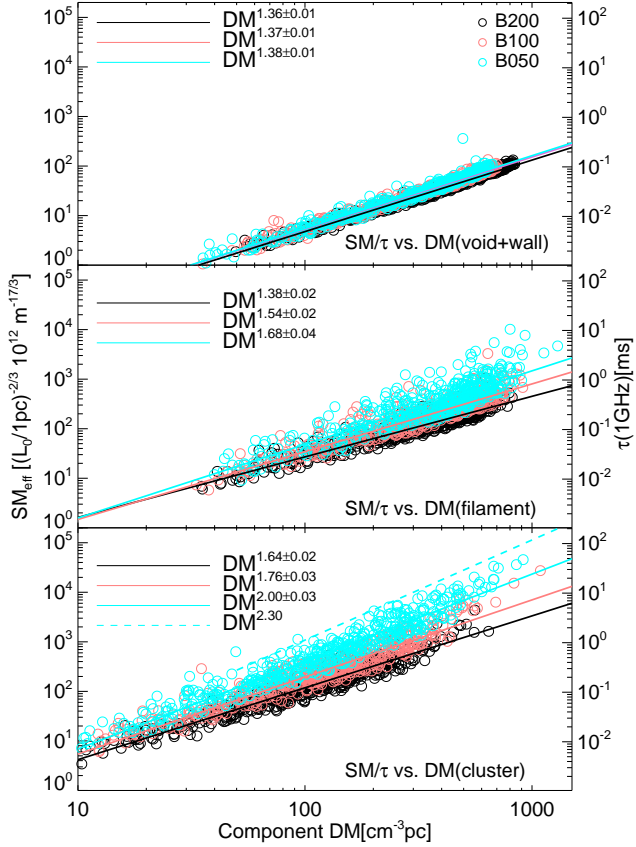}
\vspace{-0.3cm}
\caption{Top: The dispersion measure and effective scattering measure caused by the gas residing in voids and walls and the corresponding $\tau$, assuming $r_{\rm{diff}}<l_0$ and 
$L_0=0.18$ pc; the black, pink and cyan circles indicate the mock sources in simulations B200, B100 and B050, respectively. Middle (bottom): identical to the top panel but for those caused by the gas in 
filaments (clusters).}
\vspace{-0.3cm}
\label{figure7}
\end{figure}

In section 3, we find that the baryons in the clusters dominates the scattering measure and contributes approximately $15\%$ of the dispersion measure due to the IGM. Hence, the global relation between $
\tau_{\rm{IGM}}$ and $\rm{DM}_{\rm{IGM}}$ should be largely determined by the scaling of $\tau_{\rm{cluster}}-{\rm{DM_{IGM,cluster}}}$. Figure 8 presents an example based on simulation B050. This property can explain the result of the global scaling relation obtained in the last subsection. The scaling relation of $\tau \propto {\rm{DM^{2.00}}}$ is consistent with some previous analysis for 
a homogeneous turbulent scattering medium (e.g., see \citealt{Cordes2016} ). Meanwhile, it is shallower than the scaling relation of the observed Galactic pulsars at $\rm{DM}>100 \, \rm{pc\, cm^{-3}}$. 
\citet{Cordes2016} showed that the mean scattering time of the observed pulsars could be fitted by

\begin{equation}
\begin{aligned}
\widehat{\tau}(\rm{DM})=2.98\times10^{-7}\times{\rm{DM}}^{1.4}\times\\
(1+3.55\times10^{-5}\times{\rm{DM}}^{3.1}) \, {\rm{ms}}. 
\end{aligned}
\end{equation}
As there are no signs of convergence in our simulations, the dependence may become more steep if the resolution is further increased. On the other hand, the density fluctuations in the clusters are probably more homogeneous than those in the interstellar medium of the Milky Way and result in a relatively shallower scaling relation. Other physical properties, such as the thermal and ionization states, are likely different between the ISM in the Milky Way and IGM, including the baryonic medium in clusters. Hence, the scattering law of Galactic plasma may be not applicable to the scattering in the IGM.

 \begin{figure}[htbp]
\vspace{-0.3cm}
\hspace{-0.2cm}
\includegraphics[width=0.45\textwidth]{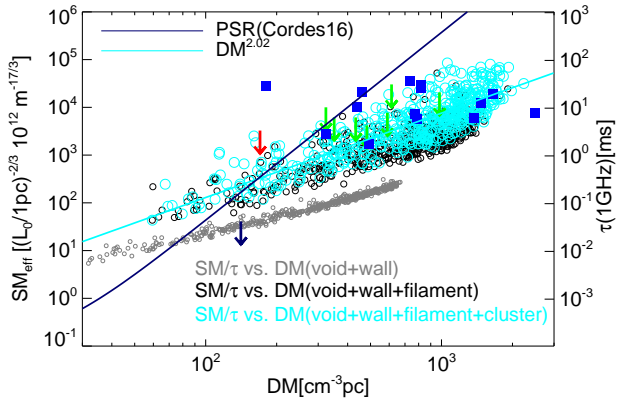}
\vspace{-0.3cm}
\caption{Effective scattering measure and the corresponding $\tau$ as a function of the dispersion measure in simulation B050 with $r_{\rm{diff}}<l_0$ and $L_0=0.18$ pc. The gray circles 
indicate the value contributed by gas in the voids and walls. The black circles indicate the value contributed by the IGM, excluding those in clusters. The cyan circles indicate the values contributed by the IGM in all  
structures.}
\vspace{-0.3cm}
\label{figure8}
\end{figure}

\subsection{Dispersion and scattering by the host galaxy and IGM}
The assumption of $\rm{DM}_{\rm{host}}=100 \, \rm{pc\, cm^{-3}}$ in section 4.1 is likely oversimplified and may overestimate the redshift of FRB events. 
\citet{Cordes2016} discussed a set of mixed models in which the dispersion and scattering of FRBs involved both the host galaxy and the IGM and suggested that the extragalactic portion of the FRBs' DM, i.e., 
$\rm{DM}_{xg}$, either consisted of mixture contributions or was dominated by the host. The scenario that the host galaxy dominated the broadening was favored in the literature, 
as the studies mainly considered the diffuse IGM, which has a relatively lower density. Our results based on simulations indicate that the gas in the filaments and clusters may play an important role, which is consistent with 
the speculation in MK13. Hence, there may be other possible solutions, e.g., the broadening of some events is dominated by the host, whereas other events are dominated by the intervention of clusters 
and filaments. 

In the upper and lower panels of Figure 9, the host's contributions to the $\rm{DM}_{xg}$ of the observed events are set to $50\%$ and $20\%$, respectively, close to one of the mixed models proposed in \citet{Cordes2016}. The cyan lines indicate the expected $\tau_{\rm{host}}$ due to $\rm{DM}_{host}$ according to the scaling law of $\tau_{\rm{host}}-{\rm{DM}_{host}}$, which was derived from 
the observed pulsars in \citealt{Cordes2016}. The circles indicate mock sources randomly distributed in the redshift ranges of $0.10<z<0.75$ and $0.10<z<1.35$ based on simulation B050. $\tau_{\rm{IGM}}$ is obtained 
according to eqn. 5a assuming $L_0=0.18\ $pc. In the top panel, the expected $\tau_{\rm{host}}$ can well explain several events in the top-left region of the $\tau-\rm{DM}$ space, 
except for FRB010724. With the fraction of the DM contributed by the host galaxy larger than $50\%$, the scaling law in \citealt{Cordes2016} can explain this event. However, the remaining events show 
an evident scattering deficit with respect to $\tau_{\rm{host}}$. Decreasing the host's contribution to $\rm{DM}_{xg}$, as in the bottom panel, can resolve the deficit. In other words, both the DM and $\tau$ 
of those events can be primarily contributed by the IGM if $L_0 \sim 0.18 \ $pc for $r_{\rm{diff}}<l_0$ or $L_0 \sim 5 \ $pc for $r_{\rm{diff}}>l_0$. Alternatively, if the host galaxy dominates the scattering while the IGM dominate the DM can also explain those events.

 \begin{figure}[htbp]
\vspace{-0.5cm}
\hspace{0.2cm}
\includegraphics[width=0.45\textwidth]{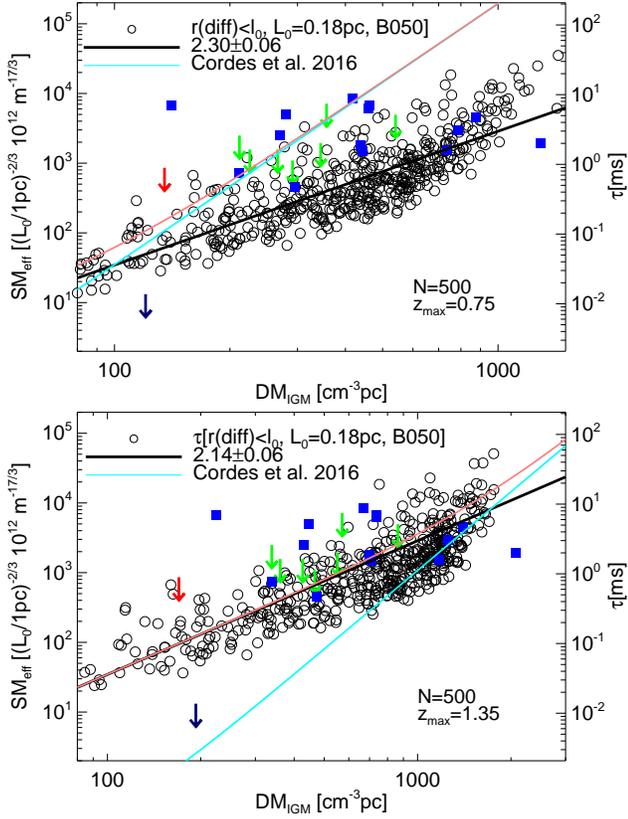}
\vspace{-0.1cm}
\caption{Same as Figure 5, but the $\rm{DM}_{IGM}$ of observed events is set to $50\%$ and $80\%$ of the $\rm{DM}_{xg}$ in the top and bottom panels, respectively. The cyan lines indicate the 
expected $\tau_{host}$ according to the scaling law given in \citet{Cordes2016}. The red lines indicate the sum of $\tau_{host}$ and fitted $\tau_{\rm{IGM}}$.}
\vspace{-0.3cm}
\label{figure9}
\end{figure}

\section{Discussion }

To make $\tau_{\rm{IGM}} \sim 5\, $ms at $\nu=1\, $GHz, the required $L_0$ here is $\sim2.5$ orders higher than the value of $10^{-3}-10^{-2} \,$pc
in the literature(\citealt{Luan2014}, XZ16). Moreover, the estimated broadening times in many theoretical works were at the frequency $\nu=0.3$ GHz. To produce $\tau_{\rm{IGM}} \sim 5\, $ms 
at $\nu=0.3\, $GHz, the required $L_0$ in B050 can be as large as 3.6 kpc if $r_{\rm{diff}}>l_0$ or 250 pc if $r_{\rm{diff}}<l_0$. Namely, the required $L_0$ can be $\sim 5$ orders
higher than the values reported in previous investigations. The latter were evaluated from the diffuse IGM in voids and walls with $\rho_b/\bar{\rho}_b \sim 1$, whereas our results are based on the 
clumpy IGM in simulation, which is mainly contributed by the gas in filaments and clusters. \cite{Luan2014} stated that the dissipation of turbulence with a notably 
small outer scale might double the temperature of the IGM on an extremely short timescale and would be incompatible with the cooling timescale, 
which is comparable to the Hubble time $t_H$. The heating timescale $\tau_{\rm{heat}}$ is approximately (XZ16)
\begin{equation}
\tau_{heat} \sim \frac{L_0}{c_s}=200(\frac{L_0}{10^{-2}\rm{pc}}) (\frac{T}{10^5\rm{K}})^{-\frac{1}{2}} \rm{yr}, 
\end{equation}
where $c_s$ is the speed of sound. If $L_0=10^{-2}$pc, $\tau_{\rm{heat}}$ is $\sim 200\, \rm{yr}$ for gas with $T=10^5$K, which is much shorter than $t_H$. The required
$L_0\sim 5\, $pc here can increase $\tau_{\rm{heat}}$ to $\sim10^5\,$yr but remains significantly shorter than $t_H$. 

This tension may be further alleviated when we consider the distribution of turbulence. The turbulence in the IGM is only well developed in the over-dense region. Hence, the turbulence 
heating is a local effect and will not change the global thermal history of the IGM, particularly the diffuse IGM in the voids and walls. Moreover, for the warm-hot ($10^5<T<10^7$K) gas in the over-dense region, 
the median ratio of turbulent kinetic energy to thermal energy was approximately $\sim 0.3$ in the adaptive mesh refined cosmological hydrodynamic simulations (e.g., \citealt{Schmidt2017}). Hence, 
turbulent heating is not expected to dramatically change the temperature. For the cold gas ($T<10^5$K) in the over-dense region, the cooling will be largely accelerated by the metal 
enrichment (\citealt{Smith2008}).

However, the required $L_0\sim 5\,$pc to produce $\tau_{\rm{IGM}} \sim 5\, $ms at $\nu=1\, $GHz is much smaller than the resolution of our simulations, and remains significantly smaller than the turbulence 
injection scale associated with cosmic structure formation, i.e., $\gtrsim 0.1$ Mpc, according to cosmological simulations(\citealt{Ryu2008}; Z13). The upper end scale of the inertial range of Kolmogorov turbulence
is usually $0.5$ dex smaller than the injection scale (\citealt{Porter1998}), which may shorten the gap to 4 orders of magnitude. In addition, we find that the effective SM and required $L_0$ increase
by a factor of $\sim 3$ when the spatial resolution of simulation increases by a factor of 2, because of improved capability to resolve over-dense region. On the other hand, increasing resolution can help to resolve 
less massive objects and their motions, and inject turbulence on scale smaller than current simulations. Thus, the gap may be significantly reduced if the simulation resolution is further increased, which, however, would 
require massive computational resources. 

While a convergence result is currently unavailable in this work, previous simulations and observations in the literature can provide some hints on the potential capability of shorten the gap by increasing resolution further. 
Figure 2 indicates that, for a fixed $L_0$, the increment on $\rm{SM_{eff}}$ in simulation with higher resolution results mainly from increased fraction of cells with baryonic density $\rho_b(z)/\bar{\rho}_b(z)>100$.  At $z=0$, the mass fraction of baryonic matter with $\rho_b/\bar{\rho}_b>100$ is $\sim 25\%$ in B050, which is smaller than a fraction of $\sim 35\% $ in \citet{Dave2010} based on a simulation with spatial resolution $\sim 5$ kpc. The mass fraction with $\rho_b/\bar{\rho}_b>1000$ is about $\sim 2 \%$ and $\sim 5\%$ at $z=1$ and $z=0$ respectively in B050,  which is smaller than $\sim 9 \%$ and $\sim 8\%$ respectively in \citet{Vogelsberger2012} based on a moving-mesh cosmological simulation with resolution of a few kpc. So far, the density fluctuation in the IGM on and below $\sim 10\, $kpc has not resolved by observations. Nevertheless, many observational efforts have been made to probe the properties of multi-phase IGM at low redshifts(e.g., \citealt{Shull2012}, \citealt{{Werk2014}}, \citealt{Danforth2016}). These studies suggested that about $30 \%$ of the cosmic baryonic matter was likely in the state of diffuse photoionized IGM with $\rho_b/\bar{\rho}_b<100$ and $T<10^5$ K, based on observation of low redshift Lyman-$\alpha$ forest.  Another $\sim 30 \%$ was in the phase of shock heated warm-hot intergalactic medium with $\rho_b/\bar{\rho}_b<100$ and $T>10^5$ K based on observation of O VI and broad Lyman absorbers. The remaining $\sim 40\%$ may reside in collapsed objects and circumgalactic gas with baryon density $\rho_b/\bar{\rho}_b>100$, and is still under investigation. In short, the mass fraction with $\rho_b/\bar{\rho}_b>100$ in the highest resolution simulation B050 in our work is lower than the results reported in the literature. The mismatch between the required $L_0$ and turbulence injection scale is expected to be alleviated by increasing resolution.

Last but not the least, if future observations find that the IGM indeed plays an important role in the scattering of FRBs, it may help to probe the gas in filaments and clusters.

\section{Conclusions}

Using cosmological hydrodynamical simulations, we investigate the dispersion and scattering of FRBs caused by the IGM. The mean value of dispersion measure 
contributed by the IGM as a function of redshift $z$ in our work is in well agreement with the literature(e.g., \citealt{McQuinn2014}). 
Moreover, we probe the contribution to $\rm{DM}_{\rm{IGM}}$ by the gas residing in various structures of the cosmic web. We find that the gas in clusters contributes approximately
$\sim 15\%-20 \%$ of the total dispersion measure caused by the IGM, the gas in filaments contributes $\sim 35-45\%$, and the gas in voids and walls contribute the remaining $\sim 35-45\%$. 
We confirm that the scattering by the clumpy IGM has significant LOS variations. We show that the scattering of FRBs by
the IGM in voids and walls is weak, but the medium in clusters and filaments can enhance the scattering by a factor of 200 in our simulation with the highest resolution. Specifically, 
the gas in clusters contributes approximately $65-80$\% of the total SM caused by the IGM, while gas in filaments contributes $20-30$\%.
We argue that the observed non-monotonic dependence of the widths of FRBs on DMs cannot determine whether the IGM is an important scattering matter of FRBs or not, considering the significant LOS variations, limited number of observed events and mixing contribution to the DMs by the host galaxy and IGM. 

Under the assumption of turbulence following Kolmogorov scaling, an 
outer scale of $\sim 5$pc is required to make $\tau_{\rm{IGM}}$ reach $\sim 1-10\,$ms at 1 GHz to explain the observed events with $\tau \geq 1\,$ms. This outer scale can 
significantly alleviate the tension regarding the timescale of turbulent dissipation and IGM cooling but remains approximately 4 orders of magnitude lower than the currently estimated turbulence 
injection scale due to structure formation. We find that the estimated effective scattering measure in our simulation is notably sensitive to the simulation resolution. With a higher resolution, 
stronger density fluctuations can be resolved. The gap in the outer scale of turbulence may be effectively shortened if the simulation resolution can be enhanced. With a mock sample of 500 
sources evenly distributed in the redshift range of $0.1<z<1.5$, the dependence of $\tau_{\rm{IGM}}$ on $\rm{DM}_{\rm{IGM}}$ is $\tau_{\rm{IGM}} \propto \rm{DM_{IGM}}^{1.56-2.02}$. The upper value 
of the scaling index, i.e., 2.02, is determined by the gas in clusters.

Factors including feedback from star formation and AGN, and supersonic turbulence may further decrease the gap in the outer scale. Feedback processes can drive the density fluctuations below tens of kpc. The density fluctuation of
supersonic turbulence differs from that of subsonic turbulence, which may change the required scales (see XZ16). Meanwhile, effects such as
 $z_L$ and $D_{\rm{eff}}$ have not been considered, which makes our results for $\tau_{{\rm{IGM}}}$ somewhat overestimated. The selection effect and intrinsic redshift 
 distribution of the FRBs remain unclear. Finally, the contributions from both the IGM and host galaxies may play important roles in the scattering of FRBs. Namely, the IGM in filaments and clusters may dominate 
 the scattering of some FRBs, whereas the host galaxy dominates others. As the number of observed events continues increasing, the dependence of $\tau$ on DM may help to ascertain the relative contributions from the host galaxy and IGM.  A more comprehensive investigation will be conducted in the future to cover these factors. 

\begin{acknowledgements}
We thank the anonymous referee for notably helpful comments that improved the manuscript. This work is supported by the National Key R\&D Program of China(2017YFB0203300), and the Key Program of the National Natural Science Foundation of China (NFSC) through grant 11733010. W.S.Z. is supported by the NSFC grant 11673077 and the Fundamental Research Funds for the Central Universities. F.L.L. is supported by the NSFC grant 11333008 and the State Key Development Program for Basic Research of China (2015CB857000). F.P.Z. is supported by the NSFC grant 11603083. 
\end{acknowledgements}

\end{document}